\documentclass[manuscript]{aastex}

\shorttitle{ChromaStarAtlas}
\shortauthors{Short \& Bayer}

\begin{document}


\title{ChromaStarAtlas: Browser-based visualization of the ATLAS9 stellar structure and spectrum grid}


\author{C. Ian Short}
\affil{Department of Astronomy \& Physics and Institute for Computational Astrophysics, Saint Mary's University,
    Halifax, NS, Canada, B3H 3C3}
\email{ian.short@smu.ca}

\author{Jason H. T. Bayer}
\affil{Department of Astronomy \& Physics and Institute for Computational Astrophysics, Saint Mary's University,
    Halifax, NS, Canada, B3H 3C3}




\begin{abstract}

ChromaStaraAtlas (CSA) is a web application that uses the ChromaStar (CS) user interface (UI) to 
allow users to navigate and display a subset of the uniformly computed comprehensive ATLAS9
grid of atmosphere and spectrum models.  It provides almost the same functionality as
the CS UI in its more basic display modes, but presents the user with primary and 
post-processed outputs, including photometric color indices, based on a properly line blanketed 
spectral energy distribution (SED).
CSA interpolates in logarithmic quantities within the subset of the ATLAS9 grid ranging
in $T_{\rm eff}$ from 3500 to 25000 K, in $\log g$ from 0.0 to 5.0, and in $[{{\rm Fe}\over {\rm H}}]$
from 0.0 to -1.0 at a fixed microturbulence parameter of 2 km s$^{-1}$, and presents outputs 
derived from the monochromatic specific intensity 
distribution, $I_\lambda$, in the $\lambda$ range from 250 to 2500 nm, and performs an
approximate continuum rectification of the corresponding flux spectrum $F_\lambda$ based on its own
internal model of the corresponding continuous extinction distribution, $\kappa^{\rm C}_\lambda$, 
based on the procedures of CS.  Optional advanced plots can be turned on that display both the
primary atmospheric structure quantities from the public ATLAS9 data files, and secondary 
structure quantities computed from internal modeling.  Unlike CS, CSA allows for activities in which
students derive $T_{\rm eff}$ values from fitting observed colors.  
The application may be found at www.ap.smu.ca/$\sim$ishort/OpenStars.     
 
\end{abstract}



\section{Introduction}

  \citet{pedagogy} described ChromaStar (CS), a general approximate atmospheric structure and emergent radiation field modeling and
post-processing code written in JavaScript (JS), the language of Web browsers, in which the visualization 
component was written in HTML.  This provides access to parameter perturbation experiments in stellar and some exoplanet
astrophysics for pedagogical purposes in way that is robustly platform-independent and suitable for 
commonplace personal and portable devices.  Because the modeling is executed on the client side, 
both the full source code and the underlying atmospheric and radiation field structures that have been computed are in 
the browser memory.  As a result, a more advanced user can optionally
display the modeled structures, and interact with the source code through the browser's developer tools while they 
are running it.  A limitation of the client-side JS modeling is that CS is limited to a line list of about twenty atomic lines and 
two TiO bands treated in the just-overlapping-line (JOLA) approximation.  Another limitation is
that CS achieves its execution speed by simply re-scaling the $T_{\rm kin}(\log\tau)$ structure with $T_{\rm eff}$ 
to obviate the need for converging thermal equilibrium, so the accuracy of the $T_{\rm kin}(\log\tau)$ structure
is limited and this in turn affects the accuracy of computed spectral features.
\citet{gss16} described ChromaStarServer (CSServ) a significant variation in which 
the CS user interface (UI) is used to interact with and visualize the results of a server-side modeling
code written in Java based on the same modeling methods as CS, but with a much larger line 
list ($\sim$ 26\, 000 lines as of this writing).  However, CSServ only draws from this line list to compute the
high resolution spectrum in a limited $\lambda$ range specified by the user, and, as with CS, the overall 
spectral energy distribution (SED) is un-blanketed.  As a result, the photometric color indices
computed with either CSServ or CS are not comparable to observed colors.  

\paragraph{}

  Here we present ChromaStarAtlas (CSA), a browser-based application for rendering and visualizing 
the uniformly computed, comprehensive grid of line-blanketed thermal equilibrium stellar atmospheric and 
surface intensity models
computed with the ATLAS9 general stellar atmospheric modeling code (\citet{castellik06}, \citet{castellik04}, 
\citet{kurucz14}).  Like CSServ, 
the atmospheric- and spectrum-related distributions are prepared on the server side in Java, and
the results sent to the client UI for some post-processing and visualization.  As of this writing,
CSA accommodates a subset of the full ATLAS9 grid consisting of models in the effective temperature 
($T_{\rm eff}$) range of 3500 to 25000 K, the
surface gravity ($\log g$) range of 0.0 to 5.0, with scaled solar abundance in the metallicity ($[{\rm Fe}/{\rm H}]$) 
range 0.0 to -1.0 with a microturbulent velocity parameter ($\xi_{\rm T}$) of 2.0 km s$^{-1}$, 
and this is a large enough region of parameter space to be useful for many important pedagogical
demonstrations and labs.  In Section \ref{sMethods} we discuss the methods used to quickly
read the ATLAS9 grid and to interpolate within it, and in Section \ref{sApps} we discuss some of the
key activities CSA allows for, with emphasis on the contrast with CS and CSServ.
The application may be found at www.ap.smu.ca/$\sim$ishort/OpenStars.     

\section{Methods \label{sMethods}}

\subsection{ATLAS9 grid preparation}

Details of the ATLAS9 model grid can be found in \citet{castellik06} and \citet{castellik04} and here we review those grid properties that
are especially relevant to the challenge of quickly reading and processing the data files. 
The ATLAS9 model grids span a range in $T_{\rm eff}$ from 3500 to 50000 K with a sampling, $\Delta T_{\rm eff}$, of 250 K 
for $T_{\rm eff} < 13000$ K and of 1000 K for $T_{\rm eff} \ge 13000$ K, and span a range in $\log g$ from as low 
as 0.0 at the cool end of the grid to 5.0 for all $T_{\rm eff}$ values, with a sampling, $\Delta\log g$ of 0.5, for a 
total of 476 $(T_{\rm eff}, \log g)$
combinations.

\paragraph{}

For scaled solar models of a given combination of $[{\rm Fe}/{\rm H}]$ and 
 $\xi_{\rm T}$, atmospheric structures for all 476 $(T_{\rm eff}, \log g)$ combinations computed with the ``new''
opacity distribution functions (OFDs) are in one ascii character-data 
file following the naming convention of the form of ap00k2odfnew.dat
(for the case of models of $[{\rm Fe}/{\rm H}] = 0.0$ and $\xi_{\rm T} = 2.0$ km s$^{-1}$), and have a size of $\sim 4.5$ Mbytes.
Each structure block includes the values of the column mass density $\bar{\rho}$ (``RHOX'') in g cms$^{-2}$ tabulated at 
72 depth points, and this serves as the independent depth variable.  Additional columns are 1) Kinetic temperature 
$T_{\rm kin}$ (``T'') in K, 
2) Total gas pressure $P_{\rm gas}$ (``P'') in dynes cms$^{-2}$, 3) Free electron number density $n_{\rm e}$ (``XNE'') in cms$^{-3}$,
4) Rosseland mean mass extinction $\kappa_{\rm Ros}$ (``ABROSS'') in cm$^{2}$ g$^{-1}$, 5) Radiative acceleration (``ACCRAD'')
in cm s$^{-2}$, 6) Turbulent velocity (``VTURB'') in cm s$^{-1}$, 7) Convective flux $F_{\rm conv}$ (``FLXCNV''), 8) Velocity of 
convective cells (``CONV'') in cm s$^{-1}$, and 9) Local sound speed (``VELSND'') cm s$^{-1}$. 
The corresponding surface specific intensity distributions, $I_\nu(\lambda, \cos\theta, \tau=0)$, 
for all 476 $(T_{\rm eff}, \log g)$ combinations for a given $([{\rm Fe}/{\rm H}], \xi_{\rm T})$ combination are in ascii 
character-data files following the naming convention of the form of
ip00k2.pck19 (for the case of models of $[{\rm Fe}/{\rm H}] = 0.0$ and $\xi_{\rm T} = 2.0$ km s$^{-1}$) and have a size of over 67 Mbytes. 
The $I_\nu$ distribution is tabulated at 1221 $\lambda$ values ranging from 9.09 to 16000 nm at 17 $\cos\theta$ 
values ranging from 1.0 (disk center) to 0.01 for a total of 20\, 757 $I_\nu$ values per model 
(we note that the $\Delta\cos\theta$ sampling is non-uniform, decreasing with decreasing $\cos\theta$, but
is not the usual Gauss-Legendre quadrature). 
The $I_\nu$ value at disk center, $I_\nu(\cos\theta=1)$, is given in erg s$^{-1}$ cm$^{-2}$ ster$^{-1}$ Hz$^{-1}$, and 
as the value of $10^5\times I_\nu/I_\nu(\cos\theta=1)$ for the subsequent 16 $\cos\theta$ points. 

\paragraph{}

\begin{figure}
\plotone{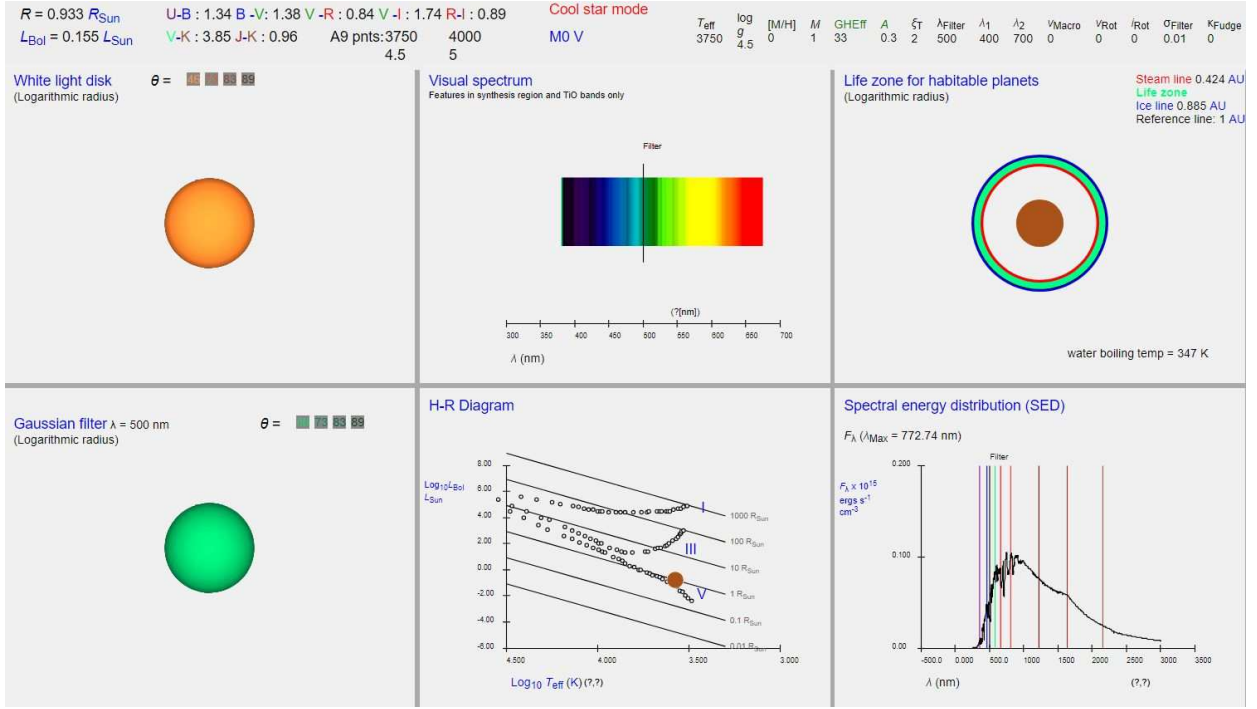}
\caption{Screenshot of the default output area for a model of ($T_{\rm eff}/\log g/[{\rm Fe}/{\rm H}]$) equal to (3750 K/4.5/0.0) 
showing the fully line blanketed emergent radiation field as i) a direct image of the flux spectrum, ii) as a plot of the SED, iii) as a white light image,
and iv) as a narrow-band image tuned to 500 nm, and v) the corresponding color indices in the Johnson $UBVRI\,JHK$ system. 
  \label{f1}
}
\end{figure}

\begin{figure}
\plotone{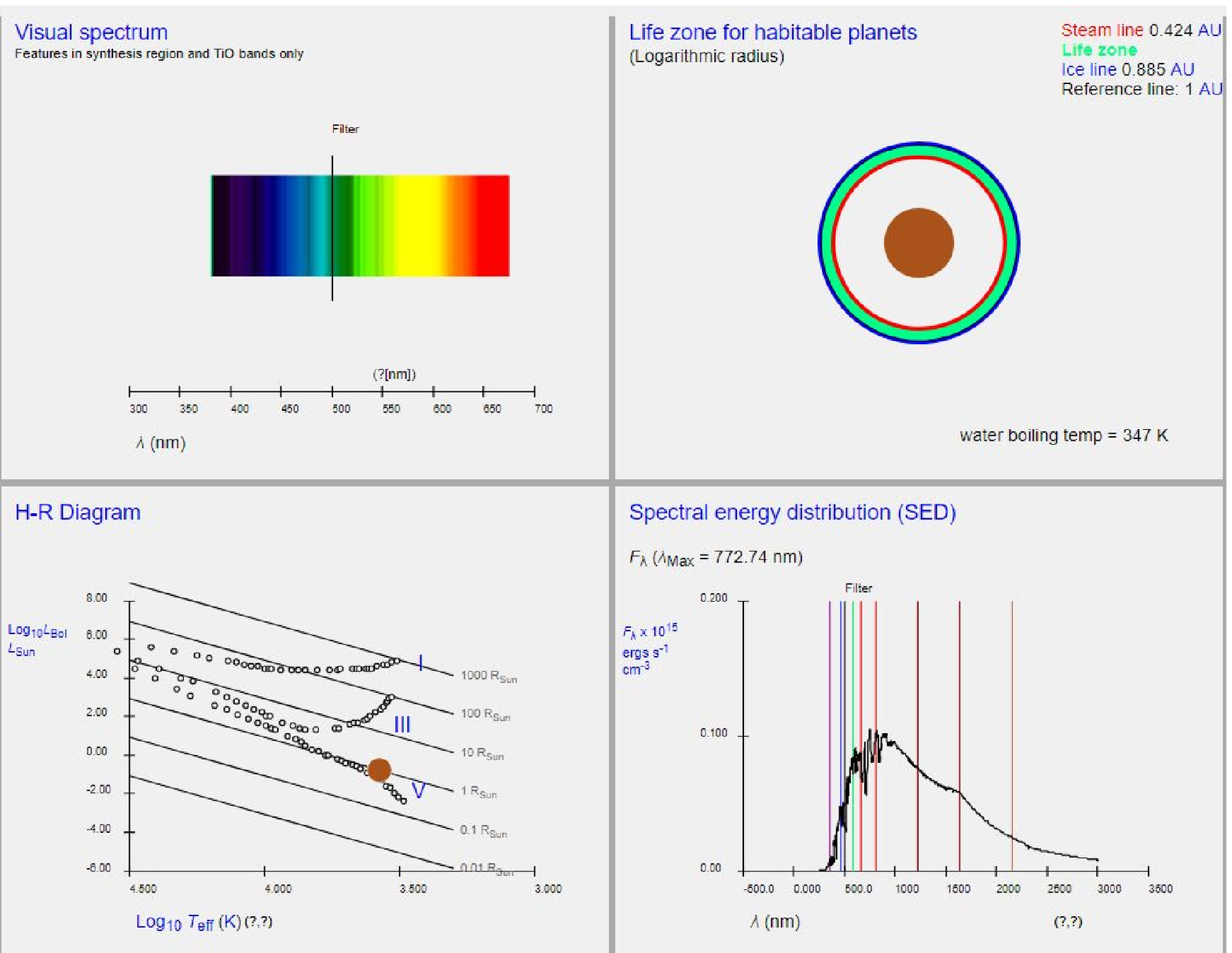}
\caption{Same as Fig. \ref{f1}, but with an expanded view of elements i) and ii).   
  \label{f2}
}
\end{figure}

  For the subset of all models of $T_{\rm eff} \le 25000$ K (406 models), we have converted the ascii structure data files for the grids of
scaled-solar $[{\rm Fe}/{\rm H}]$ value equal to 0.0, -0.5, and -1.0 and $\xi_{\rm T}$ value of 2 km s$^{-1}$
computed with the ``new'' ODFs (``ap00k2odfnew.dat'', ``am05k2odfnew.dat'', and ``am10k2odfnew.dat'') into byte data for
rapid reading by CSA.  This is performed with the auxiliary Java procedure AtlasModelServer.  The
byte-data files contain the 1) $\bar{\rho}$, 2) $T_{\rm kin}$, 3) $P_{\rm gas}$, 4) $n_{\rm e}$, 5) $\kappa_{\rm Ros}$,
and 6) $F_{\rm conv}$ structures.  The first five of these are primary atmospheric modeling quantities
that are necessary to compute other quantities of general interest, and the fifth is useful for
indicating in the visual output the depth where the atmospheric convective zone begins.
The byte-data file also contains the metallicity parameter $[{\rm Fe}/{\rm H}]$ and the logarithmic abundances, 
$\log N(Z)/N_{\rm H}$, for 100 elements 
from He ($Z=2$) to Es ($Z=99$), as specified in the ATLAS9 structure file, although many of these
for Po ($Z=84$) and higher appear to be upper limits ($\log N(Z)/N_{\rm H} = -20.0$).  
We have also pre-processed and converted the corresponding ascii $I_\nu$ distributions (files ``ip00k2.pck19'', ``im05k2.pck'', 
and ``im10k2.pck'') for all models of $T_{\rm eff} \le 25000$ K to byte data with the auxiliary Java procedure 
AtlasSpecServer.  The pre-processing includes conversion of the $10^5\times I_\nu/I_\nu(\cos\theta=1)$ values to
absolute $I_\nu$ values for all $\cos\theta$ points, conversion of the $I_\nu$ distribution to the $I_\lambda$
distribution observable with diffraction grating spectrographs, and, for all atmospheric and spectrum quantities, 
``numeric compression'' by conversion to logarithmic units.  To manage the execution time and memory 
requirements of CSA, we only include every second depth point in the structure, for a total of 36 depths, 
 and $I_\lambda$ values at a subset of the $\lambda$ values ranging from
250.50 to 2505.00 nm (608 $\lambda$ values) and at every second $\cos\theta$ point (nine $\cos\theta$ values,
including $\cos\theta=1$ and $0.01$), for a total of 5472 $\log I_\lambda$ values per model. 
Figs. \ref{f1} and \ref{f2} show a variety of renderings and signals of a model star of ($T_{\rm eff}/\log g/[{\rm Fe}/{\rm H}]$) equal to (3750 K/4.5/0.0)
that show the importance of line blanketing.

\subsection{CSA processing}

For a given set of stellar parameters input by the user, ($T_{\rm eff}$, $\log g$, and $[{\rm Fe}/{\rm H}]$),
CSA finds the two bracketing models in each of the $T_{\rm eff}$ and $\log g$ dimensions (four brackets altogether)
and interpolates the $\log\bar{\rho}$, $\log T_{\rm kin}$, $\log P_{\rm gas}$, $\log n_{\rm e}$, $\log\kappa_{\rm Ros}$, $\log F_{\rm conv}$,
and $\log I_\lambda(\lambda, \cos\theta)$ values in each of the grids of $[{\rm Fe}/{\rm H}]$ equal to 0.0 and -1.0.
A linear interpolation is first performed in $\log T_{\rm eff}$ at each of the two bracketing $\log g$ values, then in $\log g$, then in 
$[{\rm Fe}/{\rm H}]$.  It then constructs the surface flux distribution $F_\lambda(\tau=0)$ from the interpolated $I_\lambda$ 
distribution
using the $2D$ disk integration method of CSServ (see \citep{gss16}).
If a user specifies a ($T_{\rm eff}$, $\log g$, and $[{\rm Fe}/{\rm H}]$) combination that falls outside the 
subset of the ATLAS9 grid currently accommodated ({\it eg.} a low $\log g$ value at high $T_{\rm eff}$ value), then
the nearest grid model is taken.

\paragraph{}

  CSA constructs the $\log\tau_{\rm Ros}$ scale from the interpolated $\log\bar{\rho}$ and $\log\kappa_{\rm Ros}$ 
distributions, and the partial $e^-$ pressure structure $\log P_{\rm e}(\tau)$ from the interpolated $\log T_{\rm kin}$ and $\log n_{\rm e}$
structures.  It then computes the total number density of gas particles $N_{\rm gas}(\tau)$, mass density $\rho(\tau)$,
and mean molecular weight $\mu(\tau)$ structures from the ATLAS9 abundance distribution $\log N(Z)/N_{\rm H}$
scaled by the input $[{\rm Fe}/{\rm H}]$ value.  It then computes the total continuous opacity $\kappa^{\rm C}_\lambda(\tau)$
resulting 
from bound-free ({\it b-f}), free-free ({\it f-f}) and Rayleigh scattering from various species involving 
H and He, Thomson scattering from free electrons, and ground-state {\it b-f} opacity from seven metals 
(\ion{C}{1}, \ion{Mg}{1}, \ion{Mg}{2}, \ion{Al}{1}, \ion{Si}{1}, \ion{Si}{2}, and \ion{Fe}{1}) that are 
important in the UV band, and computes its own independent continuum optical depth scale $\log\tau^{\rm C}(\log\tau_{\rm Ros})$.  
It then evaluates the formal solution 
of the hydrostatic equilibrium (HSE) equation on its internal $\log\tau^{\rm C}$ scale, and computes the geometric depth scale
$z(\log\tau^{\rm C})$.
CSA then evaluates the formal solution of the radiative transfer equation (RTE) 
using its internal $\kappa^{\rm C}_\lambda$ distribution and $\log\tau^{\rm C}$ scale to compute the continuum surface intensity, 
$I^{\rm C}_\lambda(\cos\theta, \tau=0)$, and flux $F^{\rm C}_\lambda(\tau=0)$ distributions, and it uses the latter to
approximately continuum normalize (rectify) the line-blanketed $F_\lambda(\tau=0)$ spectrum.
The procedure results in an $F_\lambda(\tau=0)/F^{\rm C}_\lambda(\tau=0)$ spectrum that is 
generally continuum normalized for late-type stars to within a factor of a few around 400 nm and more accurately
around 700 nm.
The procedures for computing these secondary quantities are those as presented in \citet{gray} and are
taken from CSServ (\citet{gss16}.

\paragraph{Line labels}

   Although CSA is not performing spectrum synthesis of its own, we nevertheless read the line list as described for CSServ 
\citep{gss16} so that we can provide line identification labels in the output within the $\lambda$ range 260 to 2600 nm.  
The line list is drawn from the NIST Atomic Spectra Database \citep{nist} and contains only a modest subset of the lines in the ATLAS9 line list
and is not necessarily consistent with the line list used
to compute the spectrum, but we expect that for the relatively well-known lines that are in the NIST database the labels
will generally be relevant.

\paragraph{Limb darkening coefficients \label{sldc}}

We use the internal $I^{\rm C}_\lambda(\cos\theta)/I^{\rm C}_\lambda(\cos\theta=1)$ distribution to extract 
monochromatic continuum linear limb darkening coefficients (LDC values) following the procedure of \citet{pedagogy}.
These will reflect the $\kappa^{\rm C}_\lambda$ distribution as computed by CSA for the ATLAS9 atmospheric structure
 and not the direct ATLAS9 $\kappa^{\rm C}_\lambda$ distribution, to which we do not have access.
We note that the accuracy of these LDC values will be limited by the current restriction to fitting 
every second $\cos\theta$ value in the original ATLAS9 set (nine values rather than 17), as described above.


\section{Applications \label{sApps}}

   The most significant advantage that CSA has compared to CS and CSServ is that the computed photometric color indices in the Johnson-Bessel
$U_{\rm x}BVRI$ \citep{johnson66} and Johnson $HJK$ \citep{HJK} systems are those for a fully line blanketed $F_\lambda$ distribution.  
The seven indices ($U_{\rm x}-B_{\rm x}$, $B-V$,
$V-R$, $V-I$, $V-K$, $R-I$, and $J-K$) are calculated by CSA from its own construction of the 
$F_\lambda$ spectrum, and are calibrated on the Vega system with a single-point offset.
As a result, CSA allows for lab activities and projects such as the following.

\begin{itemize}

\item{Students could estimate the $T_{\rm eff}$ value of 
stars in the Yale Bright Star Catalogue \citet{hoffleitw91} by matching the catalogued $B-V$ color.}

\item{By varying the input 
$T_{\rm eff}$, $\log$ and mass, $M$, 
values over the full range they could also construct a model $U_{\rm x}-B_{\rm x}$ {\it vs} $B-V$ color-color diagram for
main sequence (MS) stars, and then repeat the procedure for giants to explore the dependence of the $(U_{\rm x}-B_{\rm x}) - (B-V)$
relation on luminosity class.  The stellar data presented in the appendices of an undergraduate text such as \citet{carroll} provides the input
parameters for such an activity.}

\end{itemize}

\paragraph{}

  The ATLAS9 i$\star$.pck$\star$ files necessarily contain the SED at low spectral resolution, and the high resolution 
spectrum synthesis of CSServ is not available.  Nevertheless, the user can overplot the approximately continuum-normalized 
$F_\lambda/F^{\rm C}_\lambda$ distribution prepared from the SED (see Section \ref{sMethods}) within a limited $\lambda$ 
range of their 
choice, and this plot is annotated with line identification labels drawn from the NIST Atomic Spectra Database   
(see Section \ref{sMethods}) for the species of their choice.  At the low resolution of the SED, at least the strongest
features ({\it eg.} \ion{Ca}{2} $HK$ in late-type stars) benefit from these labels.  

\paragraph{}

   The CSA UI retains most of the input and all of the output of the basic display modes of CS and CSServ.  As a result,
in addition to the above, all of the pedagogical activities described in \citet{pedagogy} and \citet{gss16} are also 
available in CSA.  We note that the direct image of the visible band spectrum and the tune-able narrow band 
filter image (see \citet{pedagogy}) now reflect fully line-blanketed $F_\lambda$ and $I_\lambda(\cos\theta)$ distributions, 
respectively, and the combination of these two outputs is useful for exploring the connection between the $I_\lambda(\lambda)$
and the $I_\lambda(\cos\theta)$ distributions.  This allows demonstration of the dependence of inferred effective angular diameter
on the wavelength of a narrow-band observation, and has relevance to exo-planet transit photometry and stellar interferometry.
Moreover, the color-coded markers on the more advanced plots indicating the connection between the $T_{\rm kin}$ structure 
and the $I_\lambda(\cos\theta)$ distribution are still provided, and provide an important aid to connecting vertical structure
to emergent observables (the LTE Eddington-Barbier relation). 

\section{Discussion \label{sDiscuss}}

   CSA represents the latest extension of our effort to fully realize the implications for education and public 
outreach (EPO) of increasingly ubiquitous computing power outside the data centers, and the radical
platform independence and instant access to that computing power made available through modern platform independent and web-aware  
programming languages.  CS and CSServe explored the potential of entirely {\it in situ} modeling in this 
environment, and CSA complements that initiative by making the much more realistic results of research-grade
modeling available in the same environment.  Now that the concept has been proven, we continue to encourage the
community to imagine the next steps.



\acknowledgments
The author acknowledges Natural Sciences and Engineering Research Council of 
Canada (NSERC) grant RGPIN-2014-03979.

\clearpage



\clearpage







\begin{thebibliography}{}


\bibitem[Castelli \& Kurucz (2006)]{castellik06} Castelli \& F. Kurucz, R. L., 2006, \aap, 454, 333
\bibitem[Castelli \& Kurucz (2004)]{castellik04} Castelli \& F. Kurucz, R. L., 2004, Proceedings of the IAU Symp. No 210, Modelling of Stellar Atmospheres, eds. N.E. Piskunov, W.W. Weiss, and D.F. Gray, poster A20 
\bibitem[Carroll \& Ostlie (2007)]{carroll} Carroll, B. W. \& Ostlie, D. A., 2007, {\it An Introduction to Modern Astrophysics}, Second Ed., Addison-Wesley
\bibitem[Gray (2005)]{gray} Gray, D.F., 2005, {\it The Observation and Analysis of Stellar Photospheres}, Third Ed., Cambridge University Press
\bibitem[Hoffleit \& Warren (1991)]{hoffleitw91}Hoffleit, E.D., Warren, Jr. W.H., ``The Bright Star Catalogue, 5th Revised Ed., 1991, Astronomical Data Center, NSSDC/ADC
\bibitem[Johnson (1965)]{HJK} Johnson, H., L., 1965, \apj, 141, 923
\bibitem[Johnson {\it et al.} (1966)]{johnson66} Johnson, H. L., Mitchell, R. I., Iriarte, B. \& Wisniewski, W. Z., 1966, Comm. Lunar Planet. Lab., 4, 99
\bibitem[Kramida {\it et al.} (2015)]{nist} Kramida, A., Ralchenko, Yu., Reader, J., and NIST ASD Team, 2015, NIST Atomic Spectra Database (ver. 5.3), [Online]. Available: http://physics.nist.gov/asd [2015, November 26]. National Institute of Standards and Technology, Gaithersburg, MD.
\bibitem[Kurucz (2014)]{kurucz14} Kurucz, R.L., 2014, Determination of Atmospheric Parameters of B-, A-, F- and G-Type Stars. Series: GeoPlanet: Earth and Planetary Sciences, Eds. E. Niemczura, B. Smalley and W. Pych, Springer International Publishing (Cham), p. 25  
\bibitem[Short (2016)]{gss16} Short, C.I., 2016, \pasp, 128, 104503, arXiv:1605.09368 
\bibitem[Short (2014)]{pedagogy} Short, C.I., 2014a, \jrasc, 108, 230, arXiv:1409.1891  
\end{thebibliography}
\end{document}